\documentclass[
aps,prl,reprint,superscriptaddress,
floatfix,longbibliography
]{revtex4-2}

\usepackage{amsmath,amssymb,amsfonts}
\usepackage{bm}
\usepackage{graphicx}
\usepackage{xcolor}
\usepackage{hyperref}
\usepackage{braket}

\begin{document}

\title{  Microscopic Bipartition and Macroscopic Symmetry in a R\'enyi Entanglement Ensemble}

\author{Zhe Wang}
\affiliation{
Department of Physics, School of Science and Research Center for
Industries of the Future, Westlake University,
Hangzhou 310030, China
}
\affiliation{
Institute of Natural Sciences, Westlake Institute for Advanced Study,
Hangzhou 310024, China
}

\author{Dongxu Liu}
\email{liudongxu@westlake.edu.cn}
\affiliation{
Department of Physics, School of Science and Research Center for
Industries of the Future, Westlake University,
Hangzhou 310030, China
}
\affiliation{
Institute of Natural Sciences, Westlake Institute for Advanced Study,
Hangzhou 310024, China
}

\author{Zheng Yan}
\email{zhengyan@westlake.edu.cn}
\affiliation{
Department of Physics, School of Science and Research Center for
Industries of the Future, Westlake University,
Hangzhou 310030, China
}
\affiliation{
Institute of Natural Sciences, Westlake Institute for Advanced Study,
Hangzhou 310024, China
}

\begin{abstract}
Universal information extracted from entanglement at a critical point is commonly
associated with the long-distance properties of the bulk state,
while microscopic details of a lattice bipartition are expected to be washed out under coarse graining.
Counterintuitively, we show directly that such microscopic information can survive in the statistical ensemble that determines the second R\'enyi entropy and obviously change the physics there.
Using quantum Monte Carlo simulations of the square-lattice
$J$-$Q_3$ model at the N\'eel--valence-bond-solid (VBS) transition,
we measure magnetic and dimer order parameters in the
R\'enyi-2 ensemble.
For a standard axis-aligned bipartition, the VBS component
perpendicular to the entanglement boundary remains finite in the
thermodynamic extrapolation and is inequivalent to the parallel
component, while the N\'eel order vanishes.
For a $45^\circ$ tilted bipartition, in contrast, the two VBS
components remain equivalent within numerical accuracy.
We interpret this contrast in terms of the R\'enyi twist defect:
the two bipartitions have the same bulk theory but realize different
microscopic defect symmetries.
Our results demonstrate that lattice information invisible to the
bulk continuum theory alone can remain active in the long-distance
R\'enyi ensemble, providing a microscopic explanation for the strong
bipartition dependence previously observed in entanglement studies
of deconfined criticality.
\end{abstract}

\maketitle

\textit{Introduction.—}
Entanglement entropy has become a central probe of universal
properties of quantum many-body states.
At $(2+1)$d quantum critical points, its subleading contributions can encode
information that is not readily accessible from conventional local
observables \cite{Calabrese2004Entanglement,Amico2008Entanglement,Laflorencie2016Quantum,Holzhey1994Geometric,Vidal2003Entanglement,Calabrese2009Entanglement,Hong2012Identifying,Fradkin2006Entanglement,Casini2007Universal,Kallin2014Corner,Bueno2015Universality,Ryu2006Holographic,Metlitski2009Entanglement,Casini2009Entanglement,Kallin2011Anomalies,Bueno2016Bounds,Helmes2014Entanglement,Helmes2016Universal}.
For example, universal subleading
contributions include angle-dependent logarithmic terms from
corners and  constant terms for suitable
smooth bipartitions, both encoding properties of the underlying
critical theory~\cite{Fradkin2006Entanglement,Casini2007Universal,Kallin2014Corner,Bueno2015Universality,Ryu2006Holographic,Metlitski2009Entanglement,Casini2009Entanglement,Kallin2011Anomalies,Bueno2016Bounds,Helmes2014Entanglement,Helmes2016Universal}.
The N\'eel--valence-bond-solid (VBS) transition in square-lattice
$J$-$Q$ models provides an important setting in which entanglement
can be used to probe the deconfined quantum critical point (DQCP)
scenario
\cite{Senthil2004Deconfined,Senthil2004Quantum,Sandvik2007Evidence,
Nahum2015Emergent,Wang2017Deconfined,Zhao2022Scaling,Song2025Evolution}.

\begin{figure}[t]
\centering
\includegraphics[width=\columnwidth]{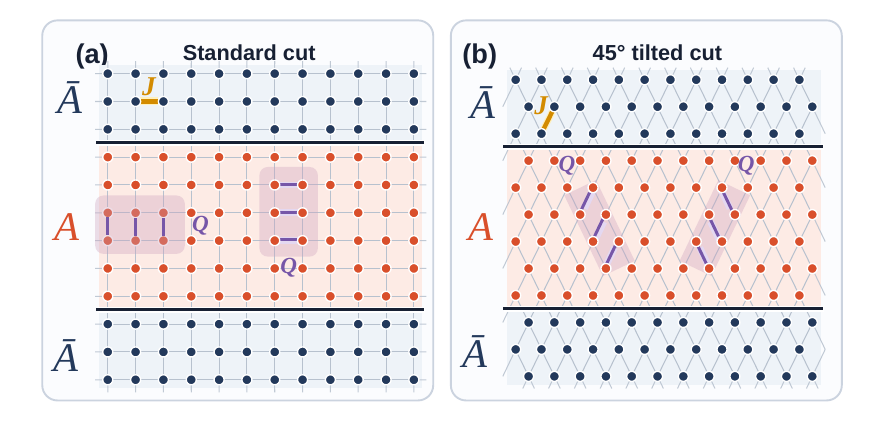}
\vspace{-10pt}
\caption{Model and Bipartitions. The red sites belong to subsystem $A$ and the dark sites to its
complement $\bar A$.
The solid lines denote the entanglement boundaries.
(a) Standard axis-aligned bipartition, for which the directions
parallel and perpendicular to the boundary are inequivalent.
(b) $45^\circ$ tilted bipartition, which treats the two columnar VBS
orientations equivalently.
The full system is periodic in both spatial directions.
}
\label{fig:model}
\vspace{-10pt}
\end{figure}

Recent R\'enyi entanglement-entropy calculations, however, have revealed a
striking sensitivity to the microscopic realization of the
bipartition \cite{Deng2024Diagnosing,DEmidio2024Entanglement,Zhao2022Scaling,Wang2026Universal}.
For a conventional axis-aligned cut [Fig.\ref{fig:model} (a)], a logarithmic contribution was
found even for a smooth entanglement boundary and was interpreted in
terms of Goldstone modes
\cite{Metlitski2011Entanglement,Kulchytskyy2015Detecting} associated with an emergent
$SO(5)\rightarrow O(4)$ symmetry breaking, consistent with a weakly
first-order scenario
\cite{Takahashi2020Valence,Deng2024Diagnosing,DEmidio2023Diagnosing,Takahashi2024SO5}.
Subsequent work showed that a $45^\circ$ tilted bipartition [Fig.\ref{fig:model} (b)] removes
the smooth-boundary logarithm \cite{DEmidio2024Entanglement} and yields corner contributions in
remarkable agreement with the prediction of an $SO(5)$-symmetric critical theory \cite{Whitsitt2017Entanglement,DEmidio2024Entanglement}.
It was further proposed that the conventional cut treats the different columnar VBS configurations unequally, whereas the tilted geometry accommodates them symmetrically.

The tilted geometry, however, is not simply a symmetry-preserving version of the same entanglement boundary.
Recent work has shown that it can also support nonordinary boundary physics, including additional gapless edge modes, whose signatures are inherited by the low-lying entanglement spectrum and can contribute to the entanglement entropy \cite{Zhu2026Bipartite}.
The strong bipartition dependence of the entropy may therefore
reflect both the microscopic symmetry of the cut and the
boundary or defect physics selected by that geometry.
Determining how these ingredients separately contribute to a
particular logarithmic entropy coefficient is an important problem,
but is not the question addressed here.

Instead, these developments expose a more fundamental issue:
\emph{What determines the long-distance symmetry structure of the
R\'enyi ensemble underlying the entanglement entropy?}
A continuum bulk theory is obtained by coarse graining the
microscopic lattice, and lattice anisotropies that are irrelevant
under bulk RG flow do not appear in the leading infrared
fixed-point description.
If the  symmetry structure of the R\'enyi ensemble
were fixed solely by this bulk theory, microscopic information about
how the entanglement boundary is embedded in the lattice would not
be expected to produce a finite macroscopic symmetry response.
Can information that is absent from the leading bulk fixed-point
theory nevertheless remain encoded in the R\'enyi ensemble after
coarse graining?

The replica construction makes clear why this question is not
determined by the bulk theory alone.
A bipartition introduces a R\'enyi twist defect at the entanglement
boundary \cite{Bianchi2016Renyi,Andrei2020Boundary}, whose symmetry
constraints and RG flow constitute additional data beyond those of
the translationally invariant bulk
\cite{Diehl1997Theory,Krishnan2023Plane,Cuomo2022Renormalization}.
The central issue is then whether this defect information can remain
confined to the entanglement boundary, or instead reorganize the
macroscopic symmetry structure of the R\'enyi ensemble itself.

Here we address this question directly by measuring N\'eel and VBS
order parameters in the statistical ensemble
$\rho_A^2/\mathrm{Tr}\rho_A^2$ that determines the second R\'enyi
entropy.
For the same bulk state, an axis-aligned bipartition produces a
finite orientation-selective VBS response: the component
perpendicular to the entanglement boundary remains finite in the
thermodynamic extrapolation and is inequivalent to the parallel
component, while the N\'eel order is consistent with zero.
For the $45^\circ$ tilted bipartition, by contrast, the two VBS
components remain equivalent within numerical accuracy.
Thus microscopic symmetry information carried by the bipartition
can remain visible after coarse graining as a macroscopic symmetry
response of the R\'enyi-2 ensemble.
Our results demonstrate that bulk infrared universality alone is not sufficient to fix the symmetry structure of a R\'enyi entanglement ensemble; the R\'enyi defect carries additional symmetry data that must be specified together with the bulk theory.

\textit{Model and method.—}We consider the spin-$1/2$ square-lattice $J$-$Q_3$ model
\cite{Sandvik2007Evidence,Lou2009Antiferromagnetic,Wang2022Scaling},
\begin{equation}
H=
-J\sum_{\langle ij\rangle}P_{ij}
-Q\sum_{\langle ijklmn\rangle}
P_{ij}P_{kl}P_{mn},
\label{eq:model}
\end{equation}
where$P_{ij}=\frac14-\mathbf S_i\cdot\mathbf S_j$
is the singlet projector on a nearest-neighbor bond.
We set $J=1$ and study the system at the N\'eel--VBS transition
coupling $Q=Q_c=1.49153(31)$
\cite{Wang2022Scaling}.
Periodic boundary conditions are imposed in both spatial directions.
The physical inverse temperature is chosen as $\beta=2L$, which is
sufficient to access the ground-state regime for the system sizes
considered here.

The two bipartitions are illustrated in Fig.~\ref{fig:model}.
Subsystem $A$ contains one half of the periodic system.
For the standard geometry, the entanglement boundaries are aligned
with a principal axis of the square lattice.
For the tilted geometry, the lattice is oriented by $45^\circ$
relative to the entanglement boundary, so that the two columnar VBS
orientations are treated equivalently by the microscopic cut.

For the normalized reduced density matrix $\rho_A$, we can define the
entanglement Hamiltonian $H_E=-\ln\rho_A $~\cite{Li2008Entanglement,Zhu2020Entanglement,Zhu2019Reconstructing,Mendes2020Measuring,Dalmonte2018Quantum,Redon2024Realizing,yang2026limits}.
Its partition function at entanglement inverse temperature
$\beta_E$ is
$
Z_E(\beta_E)
=
\mathrm{Tr}\,e^{-\beta_EH_E}
=
\mathrm{Tr}\,\rho_A^{\beta_E}.
$
Consequently,
$
S_2
=
-\ln\mathrm{Tr}\rho_A^2
=
-\ln Z_E(2).
$

More importantly for the present work, an observable
$\mathcal O$ supported in $A$ can be evaluated in the same
R\'enyi-2 ensemble according to
$
\langle\mathcal O\rangle_2
=
\frac{
\mathrm{Tr}\left(\rho_A^2\mathcal O\right)
}{
\mathrm{Tr}\rho_A^2
}.
\label{eq:renyiobs}
$
It is a thermal expectation value of
$H_E$ at the fixed entanglement inverse temperature $\beta_E=2.$ This point is important: the present study does not concern the
zero-temperature limit of $H_E$ or an extrapolation along an
entanglement-temperature phase diagram.
It directly probes the finite-entanglement-temperature ensemble that
enters the definition of $S_2$
\cite{Hastings2010Measuring,Humeniuk2012Quantum,Zhao2022Measuring}. We evaluate $\mathcal O$  using stochastic-series-expansion
quantum Monte Carlo \cite{Sandvik1999Stochastic,yan2019sweeping,yan2022global} in the replicated reduced-density-matrix
ensemble, following the multireplica approaches developed in
Refs.~\cite{Yan2023Unlocking,Li2024Relevant,Wang2025Sudden}.

Within subsystem $A$, we measure one component of the staggered
N\'eel magnetization and the two components of the columnar VBS
order. For the standard bipartition,
$m_z=N_A^{-1}\sum_{\mathbf r\in A}(-1)^{r_x+r_y}S^z_{\mathbf r}$,
$D_x=N_{b,x}^{-1}\sum_{\mathbf r\in A}(-1)^{r_x}
\mathbf S_{\mathbf r}\!\cdot\!\mathbf S_{\mathbf r+\hat{x}}$, and
$D_y=N_{b,y}^{-1}\sum_{\mathbf r\in A}(-1)^{r_y}
\mathbf S_{\mathbf r}\!\cdot\!\mathbf S_{\mathbf r+\hat{y}}$.
For the $45^\circ$ tilted bipartition,
$m_z=N_A^{-1}\sum_{\mathbf r\in A}(-1)^{r_y}S^z_{\mathbf r}$, while
each VBS component is defined by a single sum,
$D_\mu=N_{b,\mu}^{-1}\sum_{\mathbf r\in A}
\eta_\mu(\mathbf r)\,
\mathbf S_{\mathbf r}\!\cdot\!
\mathbf S_{\mathbf r+\boldsymbol{\delta}_\mu(\mathbf r)}$.
For $D_x$,
$(\eta_x,\boldsymbol{\delta}_x)=
((-1)^{r_x+\lfloor r_y/2\rfloor},\hat{x}+\hat{y})$
for odd $r_y$ and
$((-1)^{r_x+\lfloor(r_y-1)/2\rfloor},\hat{y})$
for even $r_y$; for $D_y$,
$(\eta_y,\boldsymbol{\delta}_y)=
((-1)^{r_x+\lfloor r_y/2\rfloor},\hat{y})$
for odd $r_y$ and
$((-1)^{r_x+\lfloor(r_y+1)/2\rfloor},\hat{y}-\hat{x})$
for even $r_y$.
Here $\lfloor x\rfloor$ denotes the floor function.
All bond sums include only bonds fully contained in $A$, and
$N_{b,x}$ and $N_{b,y}$ are the corresponding bond numbers.
We analyze $|m_z|$, $|D_x|$, and $|D_y|$ to avoid cancellation among
symmetry-related finite-size sectors.

\begin{figure}[t]
\centering
\includegraphics[width=\columnwidth]{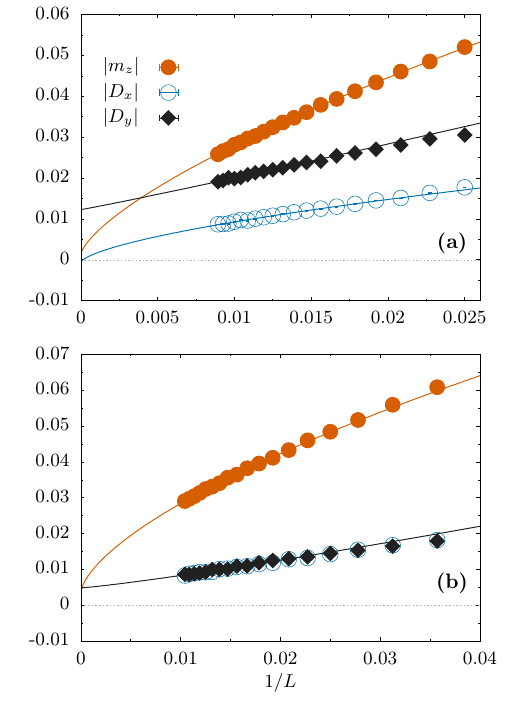}
\vspace{-30pt}
\caption{Order parameters in the R\'enyi-2 ensemble.
(a) Standard axis-aligned bipartition.
The perpendicular VBS component $|D_y|$ remains separated from
$|D_x|$ and extrapolates to a finite value, while $|m_z|$ and
$|D_x|$ are consistent with zero.
(b) $45^\circ$ tilted bipartition.
The two VBS components remain indistinguishable within statistical
accuracy over the accessible size range.
Whether their common amplitude ultimately vanishes is not required
for the symmetry comparison considered here.
Solid curves are fits to $O(L)=aL^{-k}+c$.
}
\label{fig:beta2}
\vspace{-16pt}
\end{figure}

\textit{Results.—}Figure~\ref{fig:beta2}(a) shows the finite-size dependence of the
three order parameters for the standard bipartition.
The magnetic order $|m_z|$ decreases toward zero with increasing
system size.
The VBS component $|D_x|$, parallel to the entanglement boundary in
our convention, displays a similar trend.
In sharp contrast, the perpendicular component $|D_y|$ remains
clearly separated from $|D_x|$ and extrapolates to a nonzero value.

We fit the finite-size data using, $O(L)=aL^{-k}+c$, where $a$, $k$, and $c$ are fitting parameters and $c$ gives the
thermodynamic extrapolation.
For the standard cut, varying the minimum fitting size from
$L_{\min}=56$ to $72$ gives
$c_{D_y}=0.010(3),\ 0.011(3),\ 0.012(4)$
respectively, whereas the corresponding extrapolations of
$|D_x|$ remain consistent with zero within uncertainties.
The standard R\'enyi-2 ensemble therefore retains a finite
orientation-selective VBS response in the large-system limit.

The tilted bipartition behaves qualitatively differently, as shown
in Fig.~\ref{fig:beta2}(b).
Here $|D_x|$ and $|D_y|$ are statistically indistinguishable for all
accessible system sizes.
The central conclusion does not rely on whether their common
thermodynamic-limit amplitude is exactly zero.
Indeed, the ultimate nature of the $J$-$Q_3$ transition remains
relevant to whether asymptotically small VBS order survives in the
bulk.
Our present question is different: does the entanglement
bipartition select one VBS orientation over the other?

The data give a clear distinction.
The standard R\'enyi ensemble exhibits
$
\langle|D_y|\rangle_2
\neq
\langle|D_x|\rangle_2,
$
whereas the tilted geometry satisfies, within numerical accuracy,
$
\langle|D_y|\rangle_2
\simeq
\langle|D_x|\rangle_2 .
$
Thus the existence of VBS order and the orientation selection
induced by the bipartition are logically separate questions.
The former concerns the bulk transition; the latter is a property
of the R\'enyi ensemble and is the focus of this Letter.

\begin{table}[t]
\vspace{-2pt}
\caption{
Thermodynamic-limit constants $c$ obtained from
$O(L)=aL^{-k}+c$.
For the tilted cut, $D_x$ and $D_y$ coincide within statistical
accuracy, and their average
$D=(|D_x|+|D_y|)/2$ is fitted.
}
\label{tab:fit}
\begin{ruledtabular}
\begin{tabular}{ccccc}
Cut & $L_{\min}$ & $c(|m_z|)$ & $c(|D_x|)$ & $c(|D_y|)$\\
\hline
standard & 56 & $0.001(3)$  & $0.000(3)$ & $0.010(3)$\\
standard & 68 & $-0.004(4)$ & $0.002(2)$ & $0.011(3)$\\
standard & 72 & $0.00(1)$   & $0.003(3)$ & $0.012(4)$\\
\hline
tilted & 48 & $0.001(4)$ &
\multicolumn{2}{c}{$c(|D|)=0.004(2)$}\\
tilted & 56 & $0.003(5)$ &
\multicolumn{2}{c}{$c(|D|)=0.004(2)$}\\
tilted & 64 & $0.00(3)$ &
\multicolumn{2}{c}{$c(|D|)=0.005(4)$}\\
\end{tabular}
\vspace{-10pt}
\end{ruledtabular}

\end{table}
\textit{R\'enyi-defect interpretation.—}
The result has a simple field-theoretic interpretation.
As discussed above, the observables in Fig.~\ref{fig:beta2} are
measured directly in the ensemble
$\rho_A^2/\mathrm{Tr}\rho_A^2$ at the fixed entanglement inverse
temperature $\beta_E=2$.
The relevant question is therefore whether the microscopic
bipartition leaves a symmetry imprint on this R\'enyi-2 ensemble.

In the replica representation of $S_2$, two identical copies of the
physical system are introduced, labeled by the replica index
$a=1,2$.
Across subsystem $A$, the two replicas are cyclically glued,
$1\leftrightarrow2$, whereas in the complement $\bar A$ each replica
is connected to itself.
The entanglement boundary
$\Gamma_A\equiv\partial A$ is therefore the locus where the gluing
rule changes.
After expressing this replica-gluing geometry in continuum field-theory language, the entanglement boundary $\Gamma_A$ is represented by a R\'enyi twist defect~\cite{Bianchi2016Renyi,Andrei2020Boundary,Diehl1997Theory,Krishnan2023Plane,Cuomo2022Renormalization}.
Here ``twist'' refers to the permutation of the replica labels upon
encircling the defect, rather than to a physical rotation of the spin
or VBS order parameter.

After coarse graining the microscopic lattice degrees of freedom
within each replica while keeping the $n=2$ replica-gluing structure
fixed, the continuum description consists schematically of two bulk
replica actions together with a defect contribution,
\begin{equation}
S_{\rm rep}
\simeq
S_{\rm bulk}^{(1)}
+
S_{\rm bulk}^{(2)}
+
S_{\Gamma}.
\label{eq:defact}
\end{equation}
Here $S_{\rm rep}$ denotes the coarse-grained Euclidean action of the
replicated problem,
$S_{\rm bulk}^{(a)}$ is the bulk action of replica $a=1,2$, and
$S_{\Gamma}$ denotes the effective defect sector associated with the
entanglement boundary.
Equation~(\ref{eq:defact}) is schematic rather than a microscopic
decomposition of the original Hamiltonian.
In particular, $S_{\Gamma}$ is not an additional interaction in the
physical $J$-$Q_3$ Hamiltonian.
It represents the nontrivial replica-gluing condition together with
defect-localized perturbations allowed by the microscopic symmetries
of the chosen bipartition.

This distinction is important because the symmetry constraints and RG flow of the defect need not be the same as those of the translationally invariant bulk~\cite{Diehl1997Theory,Krishnan2023Plane,Cuomo2022Renormalization}. A microscopic lattice perturbation that is irrelevant under bulk coarse graining can disappear from the long-distance bulk action while remaining symmetry allowed in the defect sector, where the corresponding defect operator generally has a different scaling dimension. Thus bulk irrelevance does not, in general, imply defect irrelevance.

For the standard axis-aligned cut, the directions parallel and
perpendicular to $\Gamma_A$ are not symmetry related.
We denote the corresponding coarse-grained local VBS fields by
$\mathcal D_{\parallel}$ and $\mathcal D_{\perp}$.
With the convention of Fig.~\ref{fig:model}, these correspond to the
$x$ and $y$ VBS orientations, respectively.
The defect is therefore allowed to couple differently to the two
orientations.
Schematically, symmetry-selective terms can take the form
\begin{equation}
S_{\Gamma}
\supset
\int_{\Gamma_A} d\ell\,
\left[
h_{\perp}\mathcal D_{\perp}(\ell)
+
\lambda_{\rm an}
\left(
\mathcal D_{\perp}^{\,2}(\ell)
-
\mathcal D_{\parallel}^{\,2}(\ell)
\right)
+\cdots
\right].
\label{eq:defectanisotropy}
\end{equation}
Here $\ell$ parametrizes the entanglement boundary.
The coefficient $h_{\perp}$ represents a possible linear VBS field
whenever the residual symmetries of the fixed bipartition do not
forbid $\mathcal D_{\perp}$ itself, whereas
$\lambda_{\rm an}$ parametrizes an orientational anisotropy between
the perpendicular and parallel VBS fluctuations.
The symbol $\supset$ indicates that the effective defect action may
contain terms of these symmetry types; Eq.~(\ref{eq:defectanisotropy})
is not intended as a microscopic identity.
Our numerical results establish that the two orientations become
inequivalent in the R\'enyi-2 ensemble, but do not require us to
determine which microscopic defect coupling is dominant.

The $45^\circ$ tilted geometry is qualitatively different.
Its microscopic symmetries interchange the two bond orientations,
$\mathcal D_x\leftrightarrow\mathcal D_y$.
Consequently, a defect perturbation that explicitly distinguishes
the two directions is forbidden.
For example,
$\mathcal D_y^2-\mathcal D_x^2$ changes sign under this interchange
and therefore cannot appear with a nonzero coefficient in a
symmetry-preserving tilted defect action.
The tilted bipartition thus does not explicitly favor one VBS
orientation over the other, consistent with the numerical
observation that $|D_x|$ and $|D_y|$ remain equivalent within
statistical accuracy.

The important result is not simply that the standard cut is anisotropic at the lattice scale. The nontrivial observation is that the corresponding orientation selection is not washed out as the system size increases. Instead, the distinction between the two VBS components remains visible in the thermodynamic extrapolation of order parameters defined over the entire subsystem \(A\). This persistence cannot be attributed to a purely local modification of a fixed-width region adjacent to the entanglement boundary, and indicates that the microscopic bipartition leaves a long-distance imprint on the statistical structure of the R\'enyi-2 ensemble.

In this sense,
\emph{emergent bulk symmetry does not by itself guarantee emergent
R\'enyi-defect symmetry}.
This statement does not imply a failure of the bulk continuum
description.
Rather, the translationally invariant bulk theory and the replica
defect carry different symmetry information, and the latter must be
specified in addition to the former.

This separation between bulk and defect behavior has close analogues
in boundary and defect critical phenomena.
A fixed bulk universality class can support ordinary, special, or
extraordinary surface critical behaviors depending on the boundary
couplings \cite{Diehl1997Theory,Ding2018Engineering,Weber2018Nonordinary},
while defects can possess RG fixed points
and scaling properties distinct from those of the translationally
invariant bulk \cite{Krishnan2023Plane,Cuomo2022Renormalization}.
More directly, recent work on $(2+1)$-dimensional $O(3)$ quantum
critical points has shown that microscopically distinct entanglement
cuts can realize different R\'enyi-defect universality classes at
fixed R\'enyi index \cite{Zhu2026Criticality}.
Related studies have also demonstrated that boundary dimerization can
drive distinct boundary RG regimes in $S_2$ without changing the bulk
critical theory \cite{Wang2026Boundary}.
These developments establish that the R\'enyi defect can possess
nontrivial structure beyond that specified by the bulk theory.
What has remained unclear, however, is whether the microscopic
symmetry of the bipartition itself can survive as a macroscopic
symmetry selection in the finite-$n$ R\'enyi ensemble.
Our results answer this question directly:
at $n=2$, the standard and tilted bipartitions produce distinct VBS
symmetry responses, revealed by the thermodynamic behavior of
$D_x$ and $D_y$.

\textit{Discussion.—}The central result of this work is that microscopic symmetry
information encoded by an entanglement bipartition can survive
coarse graining and remain visible as a macroscopic symmetry
response of the R\'enyi ensemble.
For the same bulk state, the standard axis-aligned bipartition
produces a persistent inequivalence between $|D_y|$ and $|D_x|$,
whereas the $45^\circ$ tilted bipartition preserves the equivalence
of the two VBS components within numerical accuracy.
This distinction is therefore not confined to the microscopic
geometry of the cut.

The macroscopic character of this effect is important.
The VBS order parameters are normalized over the entire subsystem
$A$, whose area scales as $L^2$, whereas a fixed-width neighborhood
of the entanglement boundary contains only $O(L)$ degrees of
freedom.
A contribution confined to such a boundary region would therefore
vanish in a subsystem-averaged observable as $L\rightarrow\infty$.
The finite thermodynamic extrapolation of the VBS anisotropy shows
instead that the bipartition leaves a long-distance imprint on the
statistical structure of the R\'enyi-2 ensemble.

This result bears directly on the recent $J$-$Q$ entanglement
controversy, while addressing a question distinct from the precise
origin of a particular logarithmic term in the entropy.
D'Emidio and Sandvik proposed that the standard cut treats different
columnar VBS configurations unequally, whereas the tilted geometry
accommodates them symmetrically \cite{DEmidio2024Entanglement}.
Recent work has further shown that the tilted geometry can support
nonordinary gapless boundary physics that is reflected in the
low-lying entanglement spectrum and can contribute to the entropy
\cite{Zhu2026Bipartite}.
Our measurements establish directly that the standard-cut
R\'enyi-2 ensemble possesses an orientation-selective VBS response,
thereby identifying a concrete symmetry-selective defect contribution
to the entanglement problem.
This conclusion is independent of the ultimate nature of the
N\'eel--VBS transition and is fully compatible with additional
defect contributions associated with nonordinary boundary modes.

The present measurements also distinguish the symmetry of the
R\'enyi ensemble from the low-lying structure of the entanglement
Hamiltonian.
Writing $H_E=-\ln\rho_A$, the second R\'enyi ensemble is
$
\rho_{E,2}
=
\frac{e^{-2H_E}}
{\mathrm{Tr}\,e^{-2H_E}}
=
\frac{\rho_A^2}
{\mathrm{Tr}\rho_A^2},
$
so that all entanglement levels contribute with weights
$e^{-2\epsilon_\alpha}$ at the fixed entanglement inverse
temperature $\beta_E=2$.
Low-lying entanglement spectra and physical edge spectra
\cite{Li2008Entanglement,Poilblanc2010Entanglement,Yan2023Unlocking,Zhu2026Bipartite} therefore
provide valuable but not, in general, complete information about
the statistical ensemble relevant to $S_2$.
Our approach probes the symmetry of that ensemble directly.

More broadly, these results clarify why a bulk long-distance theory
alone does not completely specify an entanglement measurement.
A bipartition defines a R\'enyi defect whose symmetry constraints
and RG structure need not coincide with those of the
translationally invariant bulk.
Indeed, microscopically different entanglement cuts can realize
distinct R\'enyi-defect universality classes even at fixed R\'enyi
index \cite{Zhu2026Criticality}.
Here we show directly that the microscopic \emph{symmetry} carried
by the bipartition can survive as a macroscopic order-parameter
response of the R\'enyi ensemble itself.
The appropriate long-distance description therefore requires both
the bulk theory and the symmetry and RG data of the R\'enyi defect.
In this sense, microscopic lattice information can remain visible
at long distances without implying any breakdown of the bulk
continuum description.

\textit{Conclusion.—}We have directly measured the symmetry structure of the
$n=2$ R\'enyi ensemble at the N\'eel--VBS transition of the
square-lattice $J$-$Q_3$ model.
For a standard axis-aligned bipartition, the VBS component
perpendicular to the entanglement boundary remains finite in the
thermodynamic extrapolation and is inequivalent to the parallel
component, while the N\'eel order is consistent with zero.
For a $45^\circ$ tilted bipartition, the two VBS components remain
equivalent within numerical accuracy.
This symmetry comparison is independent of whether their common VBS
amplitude in the tilted geometry ultimately vanishes or remains
finite.

The essential result is that microscopic symmetry information
encoded in the bipartition is not washed out by coarse graining:
it survives as a macroscopic symmetry response of the R\'enyi-2
ensemble.
Thus the symmetry structure of a R\'enyi entanglement ensemble is
not fixed by the bulk long-distance theory alone.
The R\'enyi defect supplies additional symmetry constraints that
must be specified together with the bulk theory.
Our results therefore establish a direct connection between
microscopic bipartition symmetry and macroscopic order-parameter
structure in a reduced-density-matrix ensemble.

\begin{acknowledgments}
\textit{\color{black}
Acknowledgments}\,---\, ZW Supported by the National Natural Science Foundation of China under Grant No.12605078.
This project is supported by the Scientific Research Project (No. WU2025B011), Feng-Ying Career Development Chair Fund and the Start-up Funding of Westlake University, and the Natural Science Foundation of China (Grant No. 12674188), and Quantum Meta Key (QuaMeK) computing. The authors thank the IT service office and the high-performance computing center of Westlake University.
\end{acknowledgments}

\end{document}